\documentclass{Rinton-P9x6}

\begin{document}
\def\nbox#1#2{\vcenter{\hrule \hbox{\vrule height#2in
\kern#1in \vrule} \hrule}}
\def\sq{\,\raise.5pt\hbox{$\nbox{.08}{.08}$}\,}         

\title{Dark Energy and Condensate Stars:\\ Casimir Energy in the Large}
\author{Pawel O. Mazur}

\address{Department of Physics and Astronomy\\
University of South Carolina\\
Columbia, SC 29208, USA\\
E-mail: mazur@mail.psc.sc.edu}

\author{Emil Mottola}

\address{Theoretical Division,
T-8, Mail Stop B285\\
Los Alamos National Laboratory\\
Los Alamos, NM 87545 USA
\\
E-mail: emil@lanl.gov\\
{\rm \bf LA-UR-04-0659}}


\maketitle

\abstracts{Vacuum fluctuations and the Casimir effect are considered in a cosmological 
setting. It is suggested that the dark energy, which recent observations suggest make
up $73\%$ of our universe, is vacuum energy due to a causal boundary
effect at the cosmological horizon. After a discussion of the similarities
and differences between material boundaries in flat spacetime and causal
horizons in general relativity, a simple model with a purely vacuum
energy de Sitter interior and Schwarzschild exterior, separated by a thin
boundary layer is outlined. The boundary layer is a quantum transition region 
which replaces the event horizons of the classical de Sitter and Schwarzschild
solutions, through which the vacuum energy changes.} 

\section{Vacuum Fluctuations and the Cosmological Term}

Vacuum fluctuations are an essential feature of quantum theory. The attractive 
force between uncharged metallic conductors in close proximity, discovered 
and discussed by Casimir more than half a century ago, is due to the vacuum 
fluctuations of the electromagnetic field in the region between the conductors. 
At first viewed perhaps as a theoretical curiosity, the Casimir effect
is now being measured with increasing accuracy and sophistication in the laboratory.\cite{Moh}
The Casimir force directly confirms the existence of quantum fluctuations and our 
methods for handling the ultraviolet divergences they generate, to obtain meaningful 
finite answers at macroscopic distance scales. Hence these Proceedings may be a good 
occasion to emphasize that the prediction and measurement of the Casimir effect is one 
of the striking successes of relativistic quantum field theory. When combined with the 
Equivalence Principle, this success gives us some confidence that we should also be able 
to treat the effects of vacuum fluctuations at macroscopic distances in a general as well 
as a special relativistic setting.

When the gravitational effects of vacuum energy are considered, we encounter one of the 
most interesting issues at the heart of any attempt to bring general relativity into
concordance with quantum principles. Some sixty years ago W. Pauli realized that the
zero-point energy of quantum fluctuations in the vacuum should contribute to the
stress-energy tensor of Einstein's theory as would an effective cosmological constant
$\Lambda > 0$, permeating all of space uniformly with a {\it negative} pressure,
$p_{\Lambda} =-\Lambda/ 8\pi G$. Such a cosmological term in Einstein's equations
leads to spacetime becoming curved on a scale of order $\Lambda^{-{1\over 2}}$.
Estimating the influence of the zero-point energy of the radiation field--the same
vacuum fluctuations responsible for the Casimir force between conductors--with an
short distance cutoff of the order of the classical electron radius, Pauli came to the 
conclusion that the radius of the universe `could not even reach to the moon.' \cite{Pau} 
In other words, the expected curvature of space from any naive estimate of
$\Lambda$ based on vacuum fluctuations at microscopic scales is 
far greater than that actually observed.

This `cosmological constant problem' has evolved in recent years from a
theoretical question of fundamental importance to one at the center of
observational cosmology as well. Observations of type Ia supernovae at moderately
large redshifts ($z\sim 0.5$ to $1$) have led to the conclusion that the expansion
of the universe is {\it accelerating}.\cite{Per} This is possible in Einstein's theory only
if the dominant energy in the universe has an effective equation of state with
$\rho + 3p <0$, {\it i.e.} assuming its energy density $\rho >0$, it must have negative
pressure. Taken at face value the observations imply that some $73\%$ of the
energy in the universe is of this hitherto undetected (dark) variety, which leads to a
{\it measured} cosmological term in Einstein's equations of 
\begin{equation}
\Lambda_{\rm meas} \simeq (0.73) {3 H_0^2\over c^2} \simeq 1.4 \times 10^{-56} {\rm cm}^{-2}
\simeq  3.8 \times 10^{-122} {c^3 \over \hbar G}\,.
\label{cosmeas}
\end{equation}
Here $H_0$ is the present value of the Hubble parameter, approximately $75 {\rm
km/sec/Mpc} \simeq 2.4 \times 10^{-18} {\rm sec}^{-1}$. The last number in
(\ref{cosmeas}) expresses the value of the cosmological dark energy inferred from the SN
Ia data in terms of Planck units, $L_{\rm pl}^{-2} = {c^3 \over \hbar G}$. If instead of
the Planck scale, we use the scale of spontaneous symmetry breaking 
in the experimentally well-tested standard model of electroweak interactions, namely 
$1 {\rm TeV}^4  \simeq 10\ {\rm cm}^{-2}$, then $\Lambda_{\rm meas}$ is still some $57$ 
orders of magnitude smaller than this much more conservative `natural' scale.  
Such an enormous mismatch of scales and gross error of estimates strongly suggest that
some basic error is being made in estimating supposedly very weak quantum effects in gravity. 

Instead of the dark energy being tied to some fixed short distance scale, as the above
estimates assume, our experience with the Casimir effect suggests a quite different possibility, 
namely that the vacuum energy is determined instead by boundary conditions of the universe in the 
large. Just as the quartically divergent vacuum energy is not what is measured in the Casimir effect, 
but only the difference of vacuum energies with and without the conducting plates present,
so too $\Lambda_{\rm meas}$ may not be related at all to any short distance cutoff,
but instead be determined {\it dynamically} by the deviation of the universe from globally flat 
Minkowski spacetime. In that case it would not be surprising to find that $\Lambda_{\rm meas}$ 
depends on the Hubble parameter $H_0$ as in (\ref{cosmeas}) for purely dimensional reasons, 
and a non-zero value of $\rho_{\Lambda}$ comparable to the closure density would 
be expected. In such a framework the dimensionless number to be explained becomes $0.73$, rather 
than $10^{-122}$ or $10^{-57}$. In these Proceedings we would like to outline a simple model
of this kind.

\section{Event Horizons}

One important clue that quantum effects in the large could alter the predictions of
classical general relativity comes from the calculation of the stress-energy tensor
of vacuum fluctuations in curved space backgrounds possessing event horizons. Two familiar 
examples of such spacetimes are the Schwarzschild metric of an uncharged non-rotating black 
hole, and the de Sitter metric, both of which can expressed in static, spherically symmetric 
coordinates in the form,
\begin{equation}
ds^2 = -f(r)\, c^2 dt^2 + {dr^2 \over h(r)} + r^2\left( d\theta^2 + \sin^2\theta\,d\phi^2\right)
\,.
\label{sphsta}
\end{equation}
For both the Schwarzchild and de Sitter cases, the functions $f=h$ and 
\begin{equation}
\begin{array}{rll}
f_{_S}(r) &= h_{_S}(r) = 1 - {r_{_S}\over r}\,,& r_{_S} = {2GM\over c^2}\,;\\
f_{_{dS}}(r) &= h_{_{dS}}(r) = 1 - {r^2\over r_{_H}^2}\,, &r_{_H} = 
\left({3\over \Lambda}\right)^{1\over 2}\,,
\end{array}
\label{SdS}
\end{equation}
\vspace{-.2cm}

\noindent
respectively. The first metric approaches flat space at large $r$ but
becomes singular at the finite radius $r_{_S}$. The second is 
regular in the interior of a spherical region but becomes singular at a finite radius $r_{_H}$, 
depending on the cosmological vacuum energy $\rho_{_{\Lambda}} = \Lambda/8\pi G$. This is the 
cosmological horizon of an observer at the origin in a universe containing $100\%$ dark energy,
with Hubble parameter $H = c\sqrt{\Lambda/3}$. In both cases the metric singularities may
be regarded are pure coordinate artifacts, in the sense that they can be 
removed entirely by making a coordinate transformation to a different set of well-behaved 
coordinates in the vicinity of the horizons. However, it is important to recognize that this 
analytic extension of spacetime by a (singular) coordinate transformation involves a physical
assumption, namely that there are no stress-energy sources or discontinuities of any kind 
at the event horizon. Even in the classical theory the hyperbolic nature of Einstein's equations 
allows for sources and/or discontinuities transmitted at the speed of light on a null hypersurface,
such as the Schwarzschild or de Sitter horizon. 

When quantum fluctuations are considered there is an additional possibility of 
discontinuities at the horizons. This is because the behavior of the renormalized
expectation value of the stress-energy tensor, $\langle \Psi\vert T_a^{\ b}\vert\Psi \rangle$
as $r \rightarrow  r_{_S}$ or $r_{_H}$ depends on the quantum state $\vert \Psi\rangle$ 
of the field theory, specified by choosing particular solutions of the wave equation which the quantum 
field satisfies. The Schwarzschild or de Sitter horizon is a characteristic surface and regular 
singular point of the wave equation $\sq \Phi = 0$ in static coordinates (\ref{sphsta}), and its 
general solution is singular there; hence so is $\langle \Psi\vert T_a^{\ b}\vert\Psi \rangle$ in the
corresponding quantum state. Since a photon with frequency $\omega$ and energy $\hbar \omega$
far from the horizon has a local energy $E_{\rm loc}=\hbar \omega f^{-{1\over 2}}$, and the stress-energy
is dimension four, its generic behavior near the event horizon is $E_{\rm loc}^4 \sim f^{-2}$.
Detailed calculations bear this out, with the stress-energy in the vacuum state defined
by absence of quanta with respect to the static time coordinate $t$ in (\ref{sphsta})
behaving like the {\it negative} of a $p=\rho/3\sim T_{\rm loc}^4$ radiation fluid at the local 
temperature, $T_{\rm loc} = \hbar c\vert f'\vert f^{-{1\over 2}}/4\pi k_{_B}$.

Although this has been known for some time,\cite{Bou,ChrFul} the attitude usually adopted is 
that the states which lead to such divergences on the horizon are pathological, and only states 
which are regular on the horizon should be considered. In Schwarzschild spacetime the Hartle-Hawking (HH) 
state and in de Sitter spacetime the Bunch-Davies (BD) state has a regular stress tensor on the 
horizon.\cite{HH,BD} These are often considered preferred `vacuum' states because they possess the maximal symmetry 
allowed by the geodesically complete background spacetime. However, the essence of an event 
horizon is that it divides spacetime into regions which are causally disconnected from each other. 
In both the Schwarzschild and de Sitter cases, the HH and BD states respectively specify that
precise quantum correlations be set up and maintained in regions of the globally extended 
spacetimes which never have been in causal contact with each other.
Despite their mathematical appeal, it is by no means clear physically why one should
restrict attention to quantum states in which exact phase correlations between regions which
have never been in any causal contact are rigorously enforced. As soon as one drops this
acausal requirement, and on the contrary restricts attention to states with correlations 
that could have been arranged by some causal process in the past, then states 
with divergent $\langle \Psi\vert T_a^{\ b}\vert\Psi \rangle$ on the horizon become perfectly 
admissable.

Again our experience with the Casimir effect suggests a physical interpretation
and resolution of these divergences. In the calculation of the local stress tensor 
$\langle \Psi\vert T_a^{\ b}\vert\Psi \rangle$ in flat spacetime with boundaries, one also 
finds divergences in the generic situation of non-conformally invariant fields and/or curved 
boundaries.\cite{DeuCan} The divergences may be traced to the hard boundary conditions imposed on modes of all 
frequencies, even the very highest. As the theory and applications of the Casimir effect have 
developed, it became clear that the material properties of the real conductors involved in the 
experiments must be taken into account. Casimir's idealized boundary conditions on the 
electromagnetic field, appropriate for a perfect conductor with infinite conductivity, have given 
way to more detailed models incorporating the conductivity response function of real metals.\cite{Moh} 
The idealized boundary conditions which led to the divergences are not to be excluded from 
consideration; they are the correct ones at low frequencies. Finite stresses due to 
vacuum fluctuations are obtained by the modification of these boundary conditions at higher 
frequencies, required by taking into account the properties of the actual material at the boundary 
in an average continuum description. At still smaller length and time scales approaching atomic 
dimensions, the approximation of a continuous or average conductivity response function will 
have to be modified, to take account the electron band structure and microscopic graininess of 
the conductors, which are composed after all of atomic constituents.

Actually, the divergences in $\langle \Psi\vert T_a^{\ b}\vert\Psi \rangle$ at the
horizon does not require that special or artificial boundary conditions be imposed.
This may be seen from the form of the wave equation in the static coordinates (\ref{sphsta}).
Separating variables by writing $\Phi = e^{-i\omega t} Y_{lm} {\psi_{\omega l}\over r}$,
we find the second order ordinary differential equation for the radial function,
$\psi_{\omega l}$,
\begin{equation}
\left[-{d^2\over dr^{*2}} + V_l\right] \psi_{\omega l} = \omega^2 \psi_{\omega l}\,,
\label{req}
\end{equation}
where we have made the usual change of radial variable from $r$ to 
$r^* \equiv \int^r\,\frac{dr} {f(r)}$,
in order to put the second derivative term into standard form, and the
scattering potential for the mode with angular quantum number $l$ is 
\begin{equation}
V_l = f \left[ \frac{1}{r} \frac{df}{dr} + \frac{l(l+1)}{r^2} \right]\,.
\label{pot}
\end{equation}
Since $f\rightarrow 0$ linearly as $r$ approaches the horizon, the variable
$r^*\rightarrow -\infty$ logarithmically, and the potential vanishes there
as well (exponentially in $r^*$). Hence the solutions of (\ref{req}) define one dimensional
scattering states on an infinite interval (in $r^*$) with {\it free} wave
boundary conditions at the horizon. The vacuum state defined by zero occupation
number with respect to these scattering states is the Boulware vacuum $\vert \Psi_B\rangle$, 
which has a divergent $\langle \Psi_B\vert T_a^{\ b}\vert\Psi_B \rangle$, behaving
like $-T_{\rm loc}^4\ {\rm diag}(-3,1,1,1) \sim f^{-2}$ on the horizon.\cite{Bou,ChrFul}
In contrast to the Casimir effect in flat spacetime with curved boundaries, this divergence 
does not arise from hard Dirichlet or Neumann boundary conditions, but from an infinite redshift
surface with free boundary conditions. Hence the properties of no ordinary material
at the boundary can remove this divergence, and the effective cutoff of horizon
divergences can arise only from new physics in the gravitational sector at ultrashort scales, 
{\it i.e.} in the structure of spacetime itself very near to the horizon.

\section{The Model}

The suggestion that a quantum phase transition may occur in the vicinity of the classical 
Schwarzschild horizon $r_{_S}$ has been made recently.\cite{CHLS} Earlier a superfluid analogy
for gravitation had been discussed by Volovik.\cite{Vol} The present authors have
suggested that this phase transition is analogous to a Bose-Einstein condensate (BEC)
phase transition, observed in very cold atomic systems.\cite{MazMot} At a phase transition 
in which the quantum ground state rearranges itself, the vacuum energy of the state also 
changes in general. Hence the interior region may have a different effective value of $\Lambda$
than the exterior region. As is now well known from discussions of vacuum energy in the 
accelerating universe, when $p_{_V} = - \rho_{_V} < 0$, then $\rho_{_V} + 3p_{_V} < 0$ 
and this behaves like an effective repulsive term in Einstein's equations. Hence, a 
positive value of $\Lambda$ in the interior serves to support the system against 
further collapse. This may be viewed as the gravitational analog of the model of an
electron which was one of the motivations of some of the original investigations of the
Casimir effect. It was found that the Casimir force on a sphere is repulsive, and therefore
cannot cancel the repulsive Coulomb self-force on a charged sphere.\cite{Boy} However a 
repulsive Casimir force with interior vacuum energy $p_{_V} = - \rho_{_V} < 0$ is exactly what 
is needed to balance the {\it attractive} force of gravity to prevent collapse to a singularity. 
The Casimir proposal to model an elementary particle such as the electron as a conducting 
spherical shell does not work as originally proposed, but an analogous model can work
for the non-singular final state of gravitational collapse.\cite{Dir} 

The model we arrive at is one with the de Sitter interior matched to a Schwarzschild
exterior, sandwiching a thin shell which straddles the region near to $r_{_H} \simeq r_{_S}$,
cutting off the divergences in $\langle \Psi\vert T_a^{\ b}\vert\Psi \rangle$ as 
$r_{_H}$ is approached from inside and $r_{_S}$ is approached from outside. This thin
shell is the boundary layer where the new physics of a quantum phase transition 
takes place. A true quantum boundary layer requires a full quantum treatment. However as a first
approximation we may treat the boundary layer in a mean field approximation in which
Einstein's eqs. continue to hold, but with an effective equation of state of the `material'
making up the layer. The fact that this boundary layer cuts off and replaces an
infinite redshift surface at the causal boundaries of the interior and exterior suggests
that the most extreme equation of state consistent with causality should play a role here,
namely the Zel'dovich equation of state $p=\rho$, where the speed of sound becomes equal to
the speed of light. This is the critical equation of state at the limit of stability for
a phase transition to a new phase with a different value of the vacuum energy. It also
arises naturally as one component of the stress-energy tensor, 
$\langle \Psi\vert T_a^{\ b}\vert\Psi \rangle = \ {\rm diag} (-\rho, p, p_{\perp}, p_{\perp})$
which satisfies the conservation equation,
\begin{equation}
\nabla_a T^a_{\ r} = {d p\over dr} + {\rho + p\over 2f} \,{d f\over dr} 
+ 2\ \frac{ p-p_{\perp}}{r}= 0\,,
\label{cons}
\end{equation}
and has three independent components in the most general static, spherically symmetric
case. It is clear from (\ref{cons}) that the three independent components can be taken
to be that with $p = \rho/3$, behaving like $f^{-2}$, $p=\rho$,
behaving like $f^{-1}$, and $p=-\rho$, behaving like $f^{0}$, reflecting the allowed
dominant and subdominant classical scaling behaviors of the stress tensor near the horizon.

In the simplest model possible we set the tangential pressure $p_{\perp} = p$ and 
consider only two independent components of the stress tensor in non-overlapping regions 
of space. In that case we have three regions, namely,
\begin{equation}
\begin{array}{clcl}
{\rm I.}\ & {\rm Interior\ (de Sitter):}\ & 0 \le r < r_1\,,\ &\rho = - p \,,\\
{\rm II.}\ & {\rm Thin\ Shell:}\ & r_1 < r < r_2\,,\ &\rho = + p\,,\\
{\rm III.}\ & {\rm Exterior\ (Schwarzschild):}\ & r_2 < r\,,\ &\rho = p = 0\,.
\end{array}
\end{equation}
Because of (\ref{cons}), $p=-\rho$ is a constant in the interior, which
becomes a patch of de Sitter space in the static coordinates (\ref{sphsta}), for
$0\le r \le r_1 < r_{_H}$. The exterior region is a patch of Schwarzschild spacetime
for $r_{_S} < r_2 \le r < \infty$. The $p= \rho/3 < 0$ component of the stress tensor
and the smooth transition that it would make possible from one region to another has 
been neglected in this simplest model. 

The location of the interfaces at $r_1$ and $r_2$ can be estimated by the behavior of the 
stress tensor near the Schwarzshild and de Sitter horizons. If $1 - r_{_S}/r_1$ is
a small parameter $\epsilon$, then the location of the outer interface occurs at an $r_1$ 
where the most divergent term in the local stress-energy $\propto M^{-4}  \epsilon^{-2}$, becomes 
large enough to affect the classical curvature $\sim M^{-2}$, {\it i.e.} for
\begin{equation}
\epsilon \sim \frac{M_{\rm pl}} {M} \simeq 10^{-38}\ \left(\frac {M_{\odot}}{M}\right)\,,
\label{epsest}
\end{equation}
where $M_{\rm pl}$ is the Planck mass $\sqrt{\hbar c/G} \simeq 2 \times 10^{-5}$ gm.
Thus $\epsilon \ll 1$ for an object of the order of a solar mass, $M=M_{\odot}$,
with a Schwarzschild radius of order of a few kilometers. If instead of a collapsed
star one considers the interior de Sitter region to be a model of cosmological dark 
energy, then the radius $r_{_H}$ is set by the Hubble scale $c H_0^{-1} \approx 10^{28}$ cm.,
and $M \approx 3 \times 10^{22} M_{\odot}$ is the order of the total mass-energy
in the visible universe. 

Since the functions $h$ and $f$ are of order $\epsilon \ll 1$ in the transition region II, 
the proper thickness of the shell is 
\begin{equation}
\ell = \int_{r_1}^{r_2}\, dr\,h^{-{1\over 2}}  \sim \epsilon^{1\over 2} r_{_S}
\sim \sqrt {L_{\rm pl} r_{_S}} \ll r_{_S}\,.
\label{thickness}
\end{equation}
Although very small, the thickness of the shell is very much larger than the Planck 
scale $L_{\rm pl} \simeq 2 \times 10^{-33}$ cm. The energy density and pressure in the shell
are of order $M^{-2}$ and far below Planckian for $M\gg M_{\rm pl}$, so that the geometry can be 
described reliably by Einstein's equations. The details of the solution in region II,
the matching at the interfaces, $r_1$ and $r_2$, and analysis of the thermodynamic
stability of the configuration have also been studied.\cite{MazMot}

\section{Conclusions}

The simplified model illustrates the general features of bulk vacuum energy arising from
a gravitational Casimir-like boundary effect with $\rho_{_V} \simeq M_{\rm pl}/L_{\rm pl} r_{_H}^2$.
The basic assumption required for a solution of this kind to exist is that gravity, 
{\it i.e.} spacetime itself, must undergo a quantum vacuum rearrangement phase transition 
in the vicinity of the horizon, $r=r_{_S}\simeq r_{_H}$, where the value of the
vacuum energy $\rho_{_V}$ can change. This cannot occur in the strictly classical Einstein 
theory of general relativity with $\Lambda$ constant. However, quantum fluctuations alter 
the low energy effective theory of gravity and through the effects of the trace anomaly the 
conformal part of the metric becomes dynamical, unlike in the Einstein theory where it is 
constrained.\cite{MMPRD} The conformal part of the metric may provide the order parameter(s) 
that would make the analogy with atomic Bose-Einstein condensation the apt one, and the 
non-singular solution we have discussed could be called a gravitational vacuum condensate `star.' 
At the edge of the `star,' the vacuum condensate disorders, due to the quantum fluctuations
of the conformal degrees of freedom in the metric. In a mean field treatment these can
be accounted for approximately by the contributions of the anomaly to the quantum 
stress-energy tensor, $\langle \Psi\vert T_a^{\ b}\vert\Psi \rangle$ near the horizons 
of the classical Schwarzschild and de Sitter geometries. A fully consistent mean field model 
would require that this interior vacuum energy density be derived from the consideration of the 
long wavelength collective modes within the cavity. 
At a finer level of resolution this continuum mean field description must give way to a more 
fundamental treatment in terms of the analogs of the atomistic degrees of freedom that make 
up the `material' in the boundary layer, and which condense into the vacuum condensate of the 
interior.\cite{Maz} A step towards such a description would be to include fluctuations
about the mean field treatment, which will give rise to time dependent dissipative
effects as well.\cite{Mot}. The details of this relaxation to flat space will determine if
a realistic model of cosmological dark energy as a finite size Casimir effect of 
the universe in the large is possible.


\end{document}